\documentstyle[12pt]{article}
\newcommand{\nc}{\newcommand}

\nc{\dbar}{\bar{\partial}}
\nc{\be}{\begin{equation}}
\nc{\ee}{\end{equation}}

\catcode`\@=11

\@addtoreset{equation}{section}

\def\@normalsize{\@setsize\normalsize{15pt}\xiipt\@xiipt
\abovedisplayskip 14pt plus3pt minus3pt%
\belowdisplayskip \abovedisplayskip
\abovedisplayshortskip  \z@ plus3pt%
\belowdisplayshortskip  7pt plus3.5pt minus0pt}
\def\small{\@setsize\small{13.6pt}\xipt\@xipt
\abovedisplayskip 13pt plus3pt minus3pt%
\belowdisplayskip \abovedisplayskip
\abovedisplayshortskip  \z@ plus3pt%
\belowdisplayshortskip  7pt plus3.5pt minus0pt
\def\@listi{\parsep 4.5pt plus 2pt minus 1pt
            \itemsep \parsep
            \topsep 9pt plus 3pt minus 3pt}}

\def\underline#1{\relax\ifmmode\@@underline#1\else
        $\@@underline{\hbox{#1}}$\relax\fi}
\@twosidetrue
\relax

\catcode`@=12

\catcode`\@=11

\def\section{\@startsection{section}{1}{\z@}{3.5ex plus 1ex minus
   .2ex}{2.3ex plus .2ex}{\large\bf}}

\def\ps@headings{\def\@oddfoot{}\def\@evenfoot{}
\def\@oddhead{\hbox{}\hfill
        \makebox[.5\textwidth]{\raggedright\ignorespaces --\thepage{}--
        \hfill }}
\def\@evenhead{\@oddhead}
\def\subsectionmark##1{\markboth{##1}{}}
}

\ps@headings

\catcode`\@=12

\relax

\def\figcap{\section*{Figure Captions\markboth
        {FIGURECAPTIONS}{FIGURECAPTIONS}}\list
        {Fig. \arabic{enumi}:\hfill}{\settowidth\labelwidth{Fig. 999:}
        \leftmargin\labelwidth
        \advance\leftmargin\labelsep\usecounter{enumi}}}
 \relax
\def\tablecap{\section*{Table Captions\markboth
        {TABLECAPTIONS}{TABLECAPTIONS}}\list
        {Table \arabic{enumi}:\hfill}{\settowidth\labelwidth{Table 999:}
        \leftmargin\labelwidth
        \advance\leftmargin\labelsep\usecounter{enumi}}}
 \relax
\def\reflist{\section*{References\markboth
        {REFLIST}{REFLIST}}\list
        {[\arabic{enumi}]\hfill}{\settowidth\labelwidth{[999]}
        \leftmargin\labelwidth
        \advance\leftmargin\labelsep\usecounter{enumi}}}
 \relax

\catcode`\@=11

\def\ps@headings{\def\@oddfoot{}\def\@evenfoot{}
\def\@oddhead{\hbox{}\hfill
        \makebox[.5\textwidth]{\raggedright\ignorespaces --\thepage{}--
        \hfill }}
\def\@evenhead{\@oddhead}
\def\subsectionmark##1{\markboth{##1}{}}
}

\ps@headings

\relax

\def\firstpage#1#2#3#4#5#6{
\begin{document}

\begin{titlepage}
\nopagebreak
\title{\begin{flushright}
       \vspace*{-1.8in}
       {\normalsize SISSA-118/96/EP} 
\end{flushright}
\vfill
{\large \bf #3}}
\author{\large #4 \\ #5}
\maketitle
\vskip -7mm
\nopagebreak
\begin{abstract}
{\noindent #6}
\end{abstract}
\vfill
\begin{flushleft}
\rule{16.1cm}{0.2mm}\\[-3mm]
\end{flushleft}
\thispagestyle{empty}
\end{titlepage}}
\newcommand{\dal}{\raisebox{0.085cm}
{\fbox{\rule{0cm}{0.07cm}\,}}}
\newcommand{\dt}{\partial_{\langle T\rangle}}
\newcommand{\dtbar}{\partial_{\langle\bar{T}\rangle}}
\newcommand{\al}{\alpha^{\prime}}
\newcommand{\mst}{M_{\scriptscriptstyle \!S}}
\newcommand{\mpl}{M_{\scriptscriptstyle \!P}}
\newcommand{\dv}{\int{\rm d}^4x\sqrt{g}}
\newcommand{\lv}{\left\langle}
\newcommand{\rv}{\right\rangle}
\newcommand{\ph}{\varphi}
\newcommand{\sbar}{\,\bar{\! S}}
\newcommand{\xbar}{\,\bar{\! X}}
\newcommand{\fbar}{\,\bar{\! F}}
\newcommand{\zbar}{\,\bar{\! Z}}
\newcommand{\tbar}{\bar{T}}
\newcommand{\ubar}{\bar{U}}
\newcommand{\ybar}{\bar{Y}}
\newcommand{\phb}{\bar{\varphi}}
\newcommand{\cm}{Commun.\ Math.\ Phys.~}
\newcommand{\pr}{Phys.\ Rev.\ D~}
\newcommand{\prl}{Phys.\ Rev.\ Lett.~}
\newcommand{\pl}{Phys.\ Lett.\ B~}
\newcommand{\ibar}{\bar{\imath}}
\newcommand{\jbar}{\bar{\jmath}}
\newcommand{\np}{Nucl.\ Phys.\ B~}
\newcommand{\e}{{\rm e}}
\newcommand{\gsi}{\,\raisebox{-0.13cm}{$\stackrel{\textstyle
>}{\textstyle\sim}$}\,}
\newcommand{\lsi}{\,\raisebox{-0.13cm}{$\stackrel{\textstyle
<}{\textstyle\sim}$}\,}
\date{}
\firstpage{95/XX}{3122}
{\large\sc Higher Derivative F-Terms in $N=2$ strings} 
{Jose F. Morales$^{b}$ and Marco Serone$^{a,b}$}
{\normalsize\sl
$^a$Istituto Nazionale di Fisica Nucleare, sez.\ di Trieste,
Italy\\[-3mm]
\normalsize\sl $^b$International School for Advanced Studies,ISAS-SISSA
I-34100 Trieste, Italy\\[-3mm]}
{We study a special class of higher derivative F-terms of the form 
$F_{g,n}W^{2g}(\Pi f)^{n}$ where W is the $N=2$ gravitational superfield 
and $\Pi$ is the chiral projector applied to a non-holomorphic function $f$
of the heterotic dilaton vector superfield. We analyze these couplings 
in the heterotic theory on $K3\times T^2$ , where it is found they 
satisfy an anomaly equation as the well studied $F_{g,0}$ terms. 
We recognize that, near a point of SU(2) enhancement, a given generating 
function of the leading singularity of the $F_{g,n}$ reproduces the free 
energy of a c=1 string at an arbitrary 
radius R. According to the $N=2$ heterotic-type II duality in 4d, we then 
study these couplings near a conifold singularity, using its local 
description in terms of intersecting D-5-branes. In this context,
it turns out that there exists, among the other states involved, a vector 
gauge field reproducing the heterotic leading singularity structure.}

\section{Introduction and Summary}
The considerable progress in the non-perturbative understanding of string
theories have established strong links between apparently different strings
as different limits of more fundamental theories \cite{ht,ew}. 
Among other dualities, the $N=2$ string in four dimensions, realized by
dual pairs constructed from Calabi-Yau and $K3\times T^2$ compactifications
of the type II and heterotic string respectively \cite{kv}, 
have received special 
interest as an intermediate step toward more realistic $N=1$ models.

The appearence of singularities
in the vector moduli space is a common fact of these $N=2$ vacua.
They are well understood as loci
where additional charged particles become massless.
While in the heterotic vacua
a detailed analysis of these singularities can be done, since the states
which become massless are elementary states, in the type II
side the conifold singularities are related to solitonic branes, charged
under R-R gauge fields, wrapped around vanishing cycles of the Calabi-Yau 
manifold \cite{andy} and our 
possibilities of study are then more limited. The Polchinski
proposal of an exact conformal field theory for these solitonic objects
\cite{pol1,pol2} as Dirichlet branes has given strength to this proposal
and also offered us a device to perform string computations.
Rather amazingly, D-branes have been used also to construct
new vacuum configurations that locally describe the conifold in terms
of intersecting five-branes \cite{bsv}, where the hypermultiplet
becoming massless is a fundamental string state, giving us the 
opportunity of performing perturbative computations.

In this paper we will exploit this picture in order to get a better
understanding of the nature of higher derivative terms in string theory.
The study of higher derivative F-terms is crucial if
one wants to establish a string-string duality beyond the effective action
level. Although in general the control of these terms is very difficult, a
certain class of them, those of the form $F_{g}(Z)W^{2g}$, with $W$ the $N=2$
gravitational superfield, have been proven
to be controlled by holomorphic anomaly equations for both
$N=2$ Type II vacua and some heterotic dual models \cite{bcov,agnt1}. 
Moreover, these $F_g$ couplings have been also related 
to topological partition functions of twisted sigma models on  
Calabi-Yau manifolds. The
singularity structure in both side were also studied and it was 
found an exciting connection with the $c=1$ string 
at the self-dual radius \cite{deb}.

After a brief section in which we review some generalities of 
higher derivative F-terms, we complete in section three the
study of the $F_g$ couplings on the heterotic side, showing 
that for an
arbitrary model $O(2,n_V)$, $n_V$ being the number of vector multiplets
(the $O(2,1)$ case was studied in \cite{agnt}
arriving to similar results) they are controlled by an holomorphic anomaly 
equation of the same kind as the one observed for type II strings
in the weak coupling limit. We 
should point out that the computation of this anomaly equation for
$N=2$ heterotic vacua is independent of any statement of duality and valid
even in the case of heterotic models which lack of a type II dual.

Section three is devoted to the study of a sequence of generalized
F-terms for the $N=2$ heterotic string. They are of the form 
$F_{g,n}W^{2g}(\Pi f)^{n}$ where $W$ is again the gravitational superfield
and $\Pi$ is the superconformal $N=2$ chiral projector acting on a general
function $f$ of the complex vector superfields the heterotic dilaton 
belongs to. We prove the existence of
an anomaly equation for this new sequence. The singularity structure near
codimension one locus of symmetry enhancement is studied and it reproduces
the free energy of a $c=1$ string at a radius R.

Section four explores the type II picture of this new sequence of
F-couplings. After providing a local D-brane description of the conifold,
we analyze different gauge fields looking for a candidate of the
corresponding dual vectormultiplet of the dilaton, gaining some insights
about the duality map. The match between the corresponding 
structure is realized only in the leading singularity limit. 
Away from this limit the computation requires much more work.
In the last section we give our conclusions.

\section{Higher Derivative F-Terms}

In this section we briefly review some important aspects of F-couplings
and construct a generalized sequence of them relevant for
the discussions below.

The so called F-terms are constructed from the F-component of a chiral 
superfield. The most extensively studied ones are of the type
\begin{equation}
I_g = F_g(X)\, W^{2g}|_{F-comp} 
\label{W2g}
\end{equation}
where $W$ is the (weight 1) Weyl superfield \footnote{For a general
discussion of $N=2$ supergravity see ref.\cite{dlv}.},
\begin{equation}
W_{\mu\nu}^{ij}=F_{\mu\nu}^{ij}-
R_{\mu\nu\lambda\rho}\,\theta^i\sigma_{\lambda\rho}\theta^j+\dots,
\label{chirW}
\end{equation}
that is anti-self-dual in its Lorentz indices and antisymmetric in the 
indices $i,j$ labeling the two supersymmetries;
$R_{\mu\nu\lambda\rho}$ is the anti-self-dual Riemann tensor, while
$F_{\mu\nu}^{ij}$ is the (anti-self-dual) graviphoton field strength.
$F_g(X)$ is an analytic
function of the $N=2$ chiral superfields $X^I$ (of Weyl weight 1):
\begin{equation}
X^I=\hat{X}^I + \frac{1}{2}\widehat{F}^{I}_{\lambda\rho}
\epsilon_{ij} \theta^i\sigma_{\lambda\rho}\theta^j + \dots ,
\label{chirX}
\end{equation}
where $\hat{X}^{I}$, $\widehat{F}^{I}_{\lambda\rho}$ are the scalar
components and the anti-self-dual vector field strengths of $X^{I}$.

The moduli dependence of these couplings (which one would naively expect to
be holomorphic ) has been proven to be controlled by an holomorphic anomaly
equation for the $N=2$ Calabi-Yau compactification of type II strings
\cite{bcov} and for the O(2,1) conjectured heterotic dual model \cite{agnt}.
The generalization to an arbitrary $O(2,n)$ heterotic model will be the
subject of the next section. It was pointed out in \cite{agnt} that
there are holomorphic ambiguities not captured by these
anomaly equations. A discussion of the holomorphic structure
is also included in that section. In both
cases, type II and heterotic vacua, the leading singularities have
been studied and are in perfect agreement with the Strominger's
interpretation of these singularities.

More general F-terms can be constructed by including superfields which
are chiral
projections of complex vector superfields. The superconformal chiral projection
$\Pi$ is a generalization of the $\bar{D^i}^2\bar{D^j}^2$ of the rigid
supersymmetry \footnote{For $N=1$ case, $\Pi$
was worked out in detail in reference \cite{top} 
which can be consulted for details}. The new sequence of F-terms are of the
type
\begin{equation}
I_{g,n} = \tilde{F}_{g,n}(X)\, (\Pi f(X,\bar{X}))^n \,
W^{2g}|_{F-comp} ,
\label{PW2g}
\end{equation}
where $f(X,\bar{X})$ is an arbitrary function of the complex vector superfields while
$\tilde{F}_{g,n}$ is an analytic one.

The study of these kind of interactions will occupy the main
part of this paper. The
relevant amplitudes
are given by a bunch of $2n$ vector fields from the dilaton multiplet beside
the usual $2g-2$ graviphotons and two gravitons in the heterotic 
string, corresponding to the following term in the effective action:
\begin{equation}
I_{g,n} = F_{g,n}F_d^{2n}\{g(R^2)(F_f^2)^{g-1} + 2g(g-1)(RF_f)^2(F_f^2)^{g-2}\}+...
\label{seff}
\end{equation}
where $R^2=R_{abcd} R^{abcd}$, $F^2 = F_{ab}F^{ab}$ and
$(RF)^2 = (R_{abcd}F^{cd})(R^{abef}F_{ef})$ and the subindices $d,f$ refer
to the vector in the dilaton supermultiplet and the graviphoton respectively.
 It is understood
that $R_{abcd}$ and $F_{f}$ represent the anti-self-dual
parts of the Riemann tensor and graviphoton, while $F_{d}$ is the self-dual part of the field
strength corresponding to the dilaton.

In the D-brane description it will be convenient to consider an
amplitude involving the term with 
$2g$ graviphotons, $2n$ field strenghts and two gauge fields sitting in the
vector multiplet of the modulus $\mu$. 
The relevant piece of the effective action is then
\begin{equation}
I_{g,n}|_F=
\partial^2_\mu F_{g,n} F_{\mu}^2 F_d^{2n} F_f^{2g} + ...
\label{seff2}
\end{equation}

\section{Holomorphic Anomaly Equations for $O(2,n)$ Case}

We study now the moduli dependence of $F_g$-terms for arbitrary
$N=2$ heterotic vacua.
More precisely, we consider $N=2$ heterotic models with rank $n+2$
where, apart from the dilaton, the scalars in vector multiplets sit in the
coset $O(2,n)/O(2)\times O(n)$ modulo discrete identifications that
define the duality group. The case $n=1$ was discussed in detail in 
\cite{agnt}, and the generalization for $n=2$ is straightforward.
Here we will show that for any $n$ (for heterotic models $n$ is bounded by 22),
the couplings $F_g W^{2g}$ exist and satisfy a recursion
relation, following from the holomorphic anomaly equation, that coincides
with the expected recursion relation from the dual type II model in the 
weak coupling limit.

As was pointed out in the last section, the $F_g$ couplings arise 
from heterotic amplitudes $A_g$'s involving two gravitons 
and $(2 g-2)$ graviphotons. In
\cite{agnt} these amplitudes were computed
at one loop level in heterotic string and the result is in
the following expression for $F_g$:
\be                                                                             
F_g = - \frac{(4\pi 
i)^{g-1}}{4\pi^2}\int                                        
\frac{d^2\tau}{\tau_2^3}\frac{\tau_2^{2g}}{\bar{\eta}^3}            
G_g (\tau,\bar{\tau}) 
\sum_{C}                              
C_c(\bar{\tau}) 
\sum_{(P_L,P_R)               
\in \bf{\Gamma_c}} (e^{K_0/2} P_L)^{2g-2} 
q^{\frac{1}{2}|P_L|^2}               
\bar{q}^{\frac{1}{2}P_R^2},                                      
\label{fnbn}    
\ee                                                                             
where $\tau$ is the Teichmuller parameter of the world-sheet torus, $q$ 
is       
$e^{2\pi i \tau}$,
$P_L$, $\bar{P}_L$ and $\vec{P}_R$ are the left and right moving momenta
sitting in an $n+2$ real dimensional lattice $\Gamma_c$
with the $O(2,n)$ inner product 
$\frac{1}{2}(P_L\bar{P}_L'+\bar{P}_L P_L') - \vec{P}_R \vec{P}_R'$.
In general the lattice of momenta corresponding to vector multiplets 
with the inner product defined above is not self-dual.   
As a result, world sheet modular invariance implies that the vectors in the 
dual lattice must appear in the spectrum. This dual lattice splits into
several conjugacy classes labelled by $\Gamma_c$. Each of these
classes are coupled to different blocks of the remaining conformal field
theory $(c=6, \bar{c}=22-n)$ and $C_c(\bar{\tau})=(-1)^F q^{L_0-c/24}
\bar{q}^{\bar{L}_0-\bar{c}/24}$ in this block. $C_0(\bar{\tau})$ has the 
expansion $C_0(\bar{\tau})=\bar{q}^{-1}(1-n_H \bar{q})$ where $n_H$
is the number of neutral massless hypermultiplets at generic point in
the $O(2,n)/(O(2)\times O(n))$ moduli space. $G_g(\tau,\bar{\tau})$ is
the normalized correlator 
$G_g(\tau,\bar{\tau})=\frac{1}{(g!)^2}
(\frac{1}{\tau_2})^{2g} \langle \prod_{i=1}^g \int d^2x_i
Z^1\dbar Z^2(x_i) \prod_{j=1}^g
\int d^2 y_j \bar{Z}^2 \dbar \bar{Z}^1 (y_j) \rangle$
where $Z^1,Z^2$ are the two complex bosons representing the four 
space-time coordinates. In \cite{agnt} the generating function for $G_g$ was
shown to be
\be
G(\lambda,\tau,\bar{\tau})\equiv  \sum_{g=1}^{\infty} \lambda^{2g} G_g 
(\tau,\bar{\tau})                 
= \left(\frac{2\pi i\lambda 
\bar{\eta}^3}{\bar{\Theta}_1    
(\lambda,\bar{\tau})}\right)^2 e^{- \frac{\pi 
\lambda^2}{\tau_2}},                      
\label{gformula1}                                                               
\ee                                                                             
which implies the following recursion relation for $G_g$:
\be
\partial_{\tau} G_g  = -\frac{i\pi}{2} \frac{1}{\tau_2^2} G_{g-1}
\label{bnanomaly}
\ee
In the following we will also need the precise form of $P_L$ for the
$O(2,n)$ case. Let us parametrize the coset $O(2,n)/O(2)\times O(n)$ 
by the n variables $T,U,y_\alpha$, $ \alpha=3,...,n$, where $T,U$ 
are respectively the
(complexified) K\"ahler class and complex structure of the two-dimensional
target space torus, and $y_\alpha$ represent the $n-2$ Wilson lines.
The classical prepotential for this model is given by
\be
F=S (T U-y^2),
\label{prep}
\ee
and the corresponding classical K\"ahler potential is 
\be
K_{tree}=-\log~ i(S-\bar{S})[(T-\bar{T})(U-\bar{U})-
                                  (y_\alpha-\bar{y_\alpha})^2].
\ee
where, as usual, $S$ is the complex scalar associated to the dilaton, with
the normalization $<S>=\frac{\theta}{\pi}+i\frac{8\pi}{g_{s}^{2}}$.
$P_L$ is just the $N=2$ central charge given as
\be
P_L=e^{K_0/2}(n_1+n_2(\bar{T}\bar{U}-\bar{y}^2)+m_1\bar{T}+m_2\bar{U}
                                        +k_\alpha\bar{y_\alpha}),
\label{pl}
\ee
and
\be
K_0\equiv -\log[(T-\bar{T})(U-\bar{U})-(y_\alpha-\bar{y_\alpha})^2].
\ee
We are ready now to find the holomorphic anomaly equation satisfied by 
the $F_{g}$'s.
Taking antiholomorphic derivatives of (\ref{fnbn}), we find that in order
to get the generalized recursion relations for arbitrary rank $n$, the
holomorphic and antiholomorphic derivatives of $P_L,\bar{P_L}$ must
be related in a highly non-trivial way through
\be
\partial_{\bar{i}} (e^{K_0/2}P_L)=A_{\bar{i}j}\partial_j 
(e^{K_0/2}\bar{P_L}) \label{system}
\ee
After some tedious algebra one can show that
this algebraic system of $n(n+2)$ equations for the $n^2$ variables
has solution, 
which is related
to the prepotential (\ref{prep}) through
\be
A_{\bar{i}j}=C_{\bar{i}\bar{l}\bar{s}}G^{\bar{l}j}
\label{amatrix}
\ee
where
\be
C_{\bar{i}\bar{j}\bar{k}}=\partial_{\bar{i}} \partial_{\bar{j}}
                                     \partial_{\bar{k}} \bar{F}
\ee
are the Yukawa couplings, and $G^{\bar{l}j}$ the inverse of the metric.

Substituting this expression in the antiholomorphic derivatives of $F_g$, one
can easily show that for $g>1$
\begin{eqnarray}
\partial_{\bar{i}} F_g &=&\frac{i}{4\pi^3}(4\pi i)^{g-1}e^K A_{\bar{i}j}
\int d^2\tau  G_g (\tau,\bar{\tau})\,\times \label{fnam}\\ 
&\partial_{\tau}&
\bigl{[}\frac{\tau_2^{2g-3}}{\bar{\eta}^3}\sum_{C}
C_{c}(\bar{\tau})\sum_{(P_L,P_R)
\in \Gamma_c} (e^{K_0/2}P_L)^{2g-4} \partial_j
(q^{\frac{1}{2}|P_L|^2} \bar{q}^{\frac{1}{2}P_R^2})\bigr{]},
\nonumber
\end{eqnarray}
where
$$
i,j=T,U,y_\alpha
$$

Now we can perform a partial integration with respect to $\tau$. The boundary
term vanishes for a generic point on the moduli space away from the 
singularities. The
only nonvanishing contribution appears when $\partial_{\tau}$ acts on
$G_g$. Using now eq.(\ref{bnanomaly}) and the expression (\ref{amatrix}) for
$A_{\bar{i}j}$ one obtains 
\be
\partial_{\bar{i}} F_g =  2\pi i e^{2K_0} C_{\bar{i}\bar{l}\bar{s}}
                                    G^{\bar{l}j}D_j F_{g-1},
\label{fnanom1}
\ee
where $D_j$ is the K\"ahler covariant derivative, i.e.
\be
D_j F_g = (\partial_j - 2(g-1)\partial_j K)F_g, 
\label{covder}
\ee
since $F_g$ has K\"ahler weight $2g-2$. Note that similar relations for 
the holomorphic anomaly equations have been found in \cite{bdv} purely
from $N=2$ supergravity arguments.

The expression (\ref{fnanom1}) coincides with the expected recursion relation 
satisfied by $F_g$ for Type II in the K\"ahler moduli limit corresponding 
to $S\rightarrow \infty$. We should point out that these holomorphic
anomalies are, like in the type II side, related to non-localities or massless
states that contribute in the boundaries of the moduli space of Riemman
surfaces that explains the origin of the recursion relations
from handle and dividing degenerations.
As was pointed out in the last section there is an
holomorphic ambiguity which is not captured by these anomaly equations. The
determination of the holomorphic structure of $F_g$ for the Type II
string is extremely difficult since it involves integration 
of the topological partition function
of the twisted Calabi-Yau sigma models over the moduli space of genus-$g$
Riemann surfaces. However, one can compare the leading singularities in 
$F_g$ near the loci. This analysis was done in \cite{agnt} for the rank 3 
case. The generalization of these results is straightforward. The local 
coordinate that goes to zero at the locus is taken 
to be $\mu = \sqrt{i/\pi}\exp(-K_{0}/2) Z$, 
where $Z$ is the central charge of the $N=2$ supersymmetry algebra. 
Again two states become massless at each locus 
$T=U,~T U=y^2$ and  $y_\alpha=0$, corresponding to the charges 
$m_1=-m_2=\pm 1$,$n_2=\pm 1$ and $k_\alpha=\pm 1$ respectively (all the 
non-specified charges are zero ). The leading coefficient reproduces
(-2) times the free enegy of the $c=1$ string \cite{deb} as we expected 
from the discussion above. This provides further support to the
Strominger's interpretation of these singularities and conclude
our study of the $F_g$ terms.

\section{$F_{g,n}$ terms in $N=2$ heterotic vacua}

We have seen previously how it is possible to construct more general
F-terms that include chiral projections of complex vector superfields.
Let us then begin the study of the $F_{g,n}$ terms defined by
(\ref{seff}) for the $N=2$ heterotic string. 
The computation generalizes the result
of \cite{agnt}, and it is very similar to that of the $F_{g}$'s, therefore
we will stress only the differences.
The relevant amplitude is given by
\begin{eqnarray}
A_{g,n} &=& \langle V_h(p_1^{+}) V_h(p_1^-) \prod_{i=1}^{g-1} 
V_f(p^{+(i)}_1) \prod_{j=1}^{g-1} V_f(p^{-(j)}_1)
\prod_{l=1}^{n} V_d(p^{+(l)}_1)
\prod_{m=1}^{n} V_d(p^{-(m)}_1)\rangle
\nonumber \\
&=& (p_1^+)^2(p_1^-)^2
\prod_{i,j=1}^{g-1} p^{+(i)}_1 p^{-(j)}_1
\prod_{l,m=1}^{n} p^{+(l)}_1 p^{-(m)}_1
(g!)^2 (n!)^2 F_{g,n}.
\label{anfn}
\end{eqnarray}
where we have used the complex notation:
\begin{eqnarray}
a_1^{\pm}&=&\frac{1}{\sqrt{2}}(a^1\pm ia^2) \nonumber \\
a_2^{\pm}&=&\frac{1}{\sqrt{2}}(a^3\pm ia^4) 
\label{not}
\end{eqnarray}
for every spacetime vector $a^{\mu}$.

We have chosen the kinematical configuration where
$p_1^+\neq0$,
$p_2^+=p_1^-=p_2^-=0$ for half of the operators and $p_1^-\neq0$,
$p_1^+=p_2^+=p_2^-=0$ for the other half. The vertex operators,
in this configuration for the anti self-dual part of the graviton
and graviphoton, and the self-dual part of the
vector superpartner of the dilaton are respectively:
\begin{eqnarray}
V_h(p_1^{\mp})&=& (\partial Z_2^{\pm} - ip_1^{\mp}\chi_1^{\pm}\chi_2^{\pm})
\dbar Z_2^{\pm} e^{ip_1^{\mp} Z_1^{\pm}}
\nonumber \\
V_f(p_1^{\mp})&=& (\partial X - ip_1^{\mp}\chi_1^{\pm}\Psi)
\dbar Z_2^{\pm} e^{ip_1^{\mp} Z_1^{\pm}}
\nonumber \\
V_d(p_1^{\mp})&=& (\partial X - ip_1^{\mp}\chi_1^{\pm}\Psi)
\dbar Z_2^{\mp} e^{ip_1^{\mp} Z_1^{\pm}} \nonumber \\
\end{eqnarray}
where $X$ is the complex coordinate of the left-moving torus $T^2$ and
$\Psi$ is its fermionic partner with $U(1)$ charge $+1$.

As in \cite{agnt}, the full amplitude can be evaluated in the odd spin
structure, where one of the operators should be inserted in the $(-1)$-ghost
picture due to the presence of a Killing spinor on the world sheet
torus and a picture changing operator must be 
inserted to take care of the world-sheet
gravitino zero mode. Taking one of the field strength in the $(-1)$ ghost
picture which comes with a $\Psi$, and the $e^{\phi}\bar{\Psi}\partial X$ part
of the picture changing operator, we soak up the zero mode for 
$\Psi,\bar{\Psi}$. From
the remaining $(2g+2n-3)$ gauge fields in the (0)-ghost picture only the
terms involving $\partial X$ survive since the $\Psi$'s cannot be contracted
with anything. Together with the $\partial X$ appearing in the picture changing
operator, they provide a total of $(2g+2n-2)$ $\partial X$'s which 
contribute only through their zero modes. 
We then take the zero momentum limit after extracting one power of momentum 
from each vertex operator. Including the left moving part, we arrive to a
similar expression as (\ref{fnbn}) substituting $g$ by $g+n$ and
$G_{g}(\tau,\bar{\tau})$ by

\begin{eqnarray}
G_{g,n}(\tau,\bar{\tau})&=&\frac{1}{(g!)^2(n!)^2}
(\frac{1}{\tau_2})^{2g+2n}\langle
\prod_{i,j=1}^g \int d^2x_i Z_1^+\dbar Z_2^+(x_i)
\int d^2 y_j Z_1^- \dbar Z_2^- (y_j) \nonumber    \\
&&\prod_{l,m=1}^n \int d^2x_l Z_1^-\dbar Z_2^+(x_l)
\int d^2 y_m Z_1^+ \dbar Z_2^- (y_m)\rangle
\end{eqnarray}

In order to evaluate these correlation functions it is
convenient to introduce the following generating function:
\begin{eqnarray}
G(\lambda,\lambda',\tau,\bar{\tau})&\equiv &
\sum_{g=1}^{\infty} \lambda^{2g}\lambda'^{2n} G_{g,n}(\tau,\bar{\tau})=
\nonumber\\
&=&\frac{\int\prod_{i=1,2} DZ_i^{\pm} 
exp (-S_0 -S_f-S_d)}{\int\prod_{i=1,2} DZ_i^{\pm} exp (-S_0)}
\label{glamda}
\end{eqnarray}
where $S_0$ is the free action\\
$S_0= \sum_{i=1,2}\frac{1}{\pi} \int
d^2 x (\partial Z_i^+ \dbar Z_i^- + \partial Z_i^- \dbar Z_i^+)$
and $S_f,S_d$ being the generators of correlators involving graviphotons and
vectorpartners of the dilaton:
$S_f=\int d^2 x \frac{\lambda}{ \tau_2} (Z_1^+\dbar Z_2^+ + Z_1^- \dbar
Z_2^- ) $,
$S_d=\int d^2 x \frac{\lambda'}{ \tau_2} (Z_1^-\dbar Z_2^+ + Z_1^+ \dbar
Z_2^- )$ respectively.

The computation of the correlators reduce then to the evaluation of this
gaussian. Using the $\zeta$-regularization, the results of the Appendix
of \cite{agnt} inmediately generalize to
\begin{equation}
G(\lambda,\lambda',\tau,\bar{\tau})=
\left(\frac{2\pi i\lambda^+ \bar{\eta}^3}{\bar{\Theta}_1
(\lambda^{+},\bar{\tau})}\right)
\left(\frac{2\pi i\lambda^- \bar{\eta}^3}{\bar{\Theta}_1
(\lambda^{-},\bar{\tau})}\right)
e^{- \frac{\pi (\lambda^2+\lambda'^2)}{\tau_2}}
\label{gformula}
\end{equation}
where $\Theta_1 (z,\tau)$ is the odd theta-function and
$\lambda^{\pm}=\lambda\pm \lambda'$.

Two interesting points can be noticed in this generalization. The first is
that again the generating function satisfies a differential equation
\be
\partial_{\tau} G(\lambda,\lambda'\tau,\bar{\tau}) = -\frac{i\pi}{2}\frac
{\lambda^2+\lambda'^2}
{\tau_2^2} G(\lambda,\lambda'\tau,\bar{\tau})
\label{ganomaly}
\ee
which define a recursion relation for the coefficients $G_{g,n}$
of the eq.(\ref{glamda})
\be
\partial_{\tau} G_{g,n}  = -\frac{i\pi}{2} \frac{1}{\tau_2^2}
(G_{g-1,n}+G_{g,n-1})
\label{bna}
\ee
Substituting (\ref{bna}) in (\ref{fnam}) and integrating by parts as before
we found that the new sequence satisfies also an
 holomorphic anomaly equation
\be
\partial_{\bar{i}} F_{g,n} =  2\pi i e^{2K_0} C_{\bar{i}\bar{l}\bar{s}}
                                    G^{\bar{l}j}D_j( F_{g-1,n}+F_{g,n-1}),
\nonumber
\ee
We see that these higher derivative terms behave much more like the
$F_g$'s, motivating our search for similar objects in the type II
side, the subject of the next section.

The second observation refers to the singularity structure of these
F-couplings near a codimension one locus of enhanced symmetry described
by a local coordinate $\mu=\sqrt{\frac{i}{\pi}}exp(-\frac{K_0}{2})P_L$
which goes to zero. After some simple algebra and rescaling of variables,
the leading term for $F_{g,n}$ is given by the coefficient of
$\lambda^{2g}\lambda'^{2n}$ in the expansion of
\be
(\mu)^{2-2g-2n}\int d\tau_2\tau_2^{2g+2n-3}
(\frac{\pi\lambda^+}{sin \pi\lambda^+})
(\frac{\pi\lambda^-}{sin \pi\lambda^-})e^{-\tau_2}
\label{c1}
\ee
We recognize in this expression the free energy of the $c=1$
string with cosmological constant $\mu$ on a radius 
$R=\lambda^-/\lambda^+$. This provide further evidence of the profound
connection of the $c=1$ string as topological description of conifold
singularities and deserves a more detailed study.

\section{ D-brane realization}

In the local description of the conifold singularity
in terms of intersecting D-five-branes constructed by \cite{bsv},
the $F_{g}$ couplings have been recently studied in \cite{gjns}, where it 
was found the expected singularity structure. This result strongly supports
the description given by \cite{bsv} for the conifold, so
we look, in this picture, for a gauge field that reproduces the behaviour of 
the new sequence of $F_{g,n}$ couplings, computed previously
in the heterotic side. 
A careful comparison of the corresponding singular structures allows us to
extract some useful information about the involved duality map.
As first step, let us review the
construction of \cite{bsv,gjns}. Basically, conifold singularities are 
mimiced by two non-parallel D-branes lying, say, in
\begin{eqnarray}
(x_4,x_5,x_8,x_9)=0\nonumber\\
(x_6,x_7,x_8,x_9)=v
\end{eqnarray}
For $v=0$ they intersect on a $3+1$ dimensional space identified with the
spacetime. The chosen configuration breaks $\frac{3}{4}$ of the supersymmetry
leading to $N=2$ SUSY in four dimensions. The conserved supercharges are
given by \cite{pol3} the intersection of
\begin{equation}
Q_A+(\gamma^{6}\gamma^{7}\gamma^{8}\gamma^{9})_A^B\tilde{Q}_B\nonumber
\end{equation}
and
\begin{equation}
Q_A+(\gamma^{4}\gamma^{5}\gamma^{8}\gamma^{9})_A^B\tilde{Q}_B
\end{equation}
where $A,B=1,......16$ and $Q,\tilde{Q}$ are the left and right charges
respectively. More explicitly, they are  
$$ Q_{1}^{\alpha} + {\tilde Q}^{\alpha}_{1};~~~{\rm and}~~~
Q_{2}^{\alpha} - {\tilde Q}^{\alpha}_{2}.$$
together with the corresponding dotted generators, of course,
where
\be Q^{\alpha}_{k} \equiv \int e^{-\frac{1}{2}\phi}S^{\alpha}\Sigma_{k} 
\ \ \ \ \ k=1,2 \ee
and $\phi$ is the bosonization of superghost. $S^{\alpha}$ and $\Sigma_i$
are the
spin fields of the four-dimensional space-time and the internal parts 
respectively, 
and they can be given explicitly in the bosonized form as follows:
$$
S^{\pm} = e^{\pm \frac{i}{2}(\phi_1 +\phi_2)}.
$$
$$
\Sigma_k = e^{i(\frac{3}{2}-k)(\phi_3 +\phi_4) + i\frac{1}{2}\phi_5}
$$
Beside the universal and closed sector we include
 different open sectors related to the possible
boundary conditions on endpoints of the string living in the D-brane as
N-N, N-D, D-N, D-D according to which boundary condition, Neumann or 
Dirichlet, is chosen. 
The conifold singularity is then resolved by a string stretched between
the two D-branes, that give rise to a massless hypermultiplet, 
charged under the difference
of the $U(1)$ gauge fields living in the boundary.\footnote{It was argued
in \cite{gjns} that only this combination remains massless, while the sum 
adquires mass via the Cremmer-Scherk mechanism}
The analysis of the holomorphic structure of higher derivative
couplings can be done through an annulus computation of a bunch of 
graviphotons and gauge fields as defined by (\ref{seff2}).

We shall see that there is a natural
candidate vector reproducing the given behaviour near the conifold.
The computation involve a bunch of $2n$ of these vectors besides
2g graviphotons and two boundary gauge fields. The vertex operators
in the (-1)-ghost picture and in the kinematical configuration 
(in the notation (\ref{not}) ) where
half of the operators have $p_1^{+}\neq0,p_1^{-}=p_2^{\pm}=0$ 
and the rest $p_1^{-}\neq0,p_1^{+}=p_2^{\pm}=0$, are given by:
\begin{eqnarray}
V_f(p_1^\mp)&=&[\bigl(e^{-\phi}\psi^{+}_{5}\bigr)
\bigl( \dbar {Z}^{\pm}_{2} + ip_{1}^{\mp}{\tilde \psi}_{1}^{\pm}
{\tilde \psi}^{\pm}_{2}\bigr)-
\bigl( \partial Z^{\pm}_{2} + ip_{1}^{\mp}\psi_{1}^{\pm}\psi^{\pm}_{2}\bigr)
\bigl(e^{-{\tilde\phi}}{\tilde\psi}^{\pm}_{5}\bigr)\nonumber\\
&+&ip_{1}^{\mp}e^{-{1/2}{\phi}}S^{\pm}\Sigma_1e^{-{1/2}{\tilde{\phi}}}\tilde{
S}^{\pm}\tilde{\Sigma}_2
-ip_{1}^{\mp}e^{-{1/2}{\phi}}S^{\pm}\Sigma_2e^{-{1/2}{\tilde{\phi}}}\tilde{
S}^{\pm}\tilde{\Sigma}_1] e^{ip_{1}^{\mp}Z^{\pm}_{1}}.\nonumber\\
V_d(p_1^\mp)&=&[\bigl(e^{-\phi}\psi^{+}_{5}\bigr)
\bigl( \dbar {Z}^{\mp}_{2} + ip_{1}^{\mp}{\tilde \psi}_{1}^{\pm}
{\tilde \psi}^{\mp}_{2}\bigr)-
\bigl( \partial Z^{\mp}_{2} + ip_{1}^{\mp}\psi_{1}^{\pm}\psi^{\mp}_{2}\bigr)
\bigl(e^{-{\tilde\phi}}{\tilde\psi}^{\pm}_{5}\bigr)]
e^{ip_{1}^{\mp} Z^{\pm}_{1}} \\
V_b(p) &=& \int (\partial_{\tau} X^{\mu} +ip\psi
\gamma_{\tau}\psi^{\mu})e^{ip.X} \nonumber 
\label{vo1}
\end{eqnarray}
The computation follows the same steps as for the $F_g$ case considered in 
\cite{gjns}, reference that can be consulted for a more detailed analysis.
In order to cancel the total superghost charge we should include $2g+2n$ 
picture changing operators whose
only non-vanishing contribution come from $e^{-\phi}\psi_5^-\partial Z_5^+$.
Note that there is no $Z_5^-$ in the correlation and then the $Z_5^+$ contributes
only through its zero mode
$$
Z_5^+=\bar{\mu}\sigma+....
$$
providing an overall $\bar{\mu}^{2g+2n}$. This is the expected behaviour
for the leading term of $\partial^2_{\mu}F_{g,n}$ which after
integration lead to the universal $\mu^{2-2g-2n}$ as in (\ref{c1}).
The superghosts contribution cancels exactly the fermion one in the 
D-D direction; so we are left
 with a sum over the spin structures involving the remaining four directions,
which defines a triality map on the $SO(8)$ weight lattice through the
Riemmann identity:
\be
\Sigma_{\alpha}(-)^{\alpha_1\alpha_2}\theta_\alpha(x_1) \theta_\alpha(x_3)
\theta_\alpha(x_3) \theta_\alpha(x_4) =
\theta_1(x_1') \theta_1(x_2') \theta_1(x_3') \theta_1(x_4')
\nonumber
\ee
where $(\alpha_1,\alpha_2)$ are the characters of the theta functions and
\begin{eqnarray}
x_1'&=&\frac{1}{2}(x_1+x_2+x_3+x_4)\nonumber\\
x_2'&=&\frac{1}{2}(x_1+x_2-x_3-x_4)\nonumber\\
x_3'&=&\frac{1}{2}(x_1-x_2+x_3-x_4)\nonumber\\
x_4'&=&\frac{1}{2}(-x_1+x_2+x_3-x_4)\nonumber
\end{eqnarray}
are the transformed arguments.

Under this
transformation the operators (\ref{vo1}) get mapped to
\begin{eqnarray}
 V_f(p_1^{\mp}) &\rightarrow& \bigl{(} \partial_{\tau} Z_2^{\pm}
+ip_1^{\mp}(\psi_1^{\pm}-  
\tilde{\psi}_1^{\pm})(\psi_2^{\pm}- \tilde{\psi}_2^{\pm})\bigr{)} 
e^{ip_1^{\mp}Z_1^{\pm}} \nonumber\\
V_d(p_1^{\mp}) &\rightarrow& \bigl{(} \partial_{\tau} Z_2^{\mp}
+ip_1^{\mp}(\psi_3^{\pm}\psi_4^{\mp}-
\tilde{\psi}_3^{\pm}\tilde{\psi}_4^{\mp})\bigr{)}
e^{ip_1^{\mp}Z_1^{\pm}}
\end{eqnarray}
while the boundary operator goes to itself. The correlations of the bosonic
part give total derivatives which upon partial integration brings down one
power of momentum for each vertex, matching the momentum structure of
(\ref{anfn}). We are left with correlators generated by a same kind of 
generating function (\ref{glamda}) where now
$$S_f=\int  d^2 x \frac{\lambda}{t} (Z_1^+\dbar_{\tau} Z_2^+ 
+ Z_1^- \dbar_{\tau} Z_2^- ) \nonumber\\$$ and
$$S_d=\int  d^2 x \frac{\lambda'}{t} (Z_1^-\dbar_{\tau} Z_2^+ 
+ Z_1^+ \dbar_{\tau} Z_2^- )$$
where $t$ is the modulus of the world-sheet annulus.
We observe that this perturbed action has four zero modes
given by $\tilde{\psi}_1^{\pm}= \psi_1^{\pm}
=$ constant and $\tilde{\psi}_2^{\pm}= \psi_2^{\pm}
=$ constant, which should be soaked by the bilinear fermion part of the
inserted boundary gauge fields. Without this insertion the correlation
would vanish as we expected from the absence of the corresponding term
in the effective action (\ref{seff}). The
mode expansion for
the involved fields are:
\begin{eqnarray}
\psi_{1,2}^{\pm}&=& \sum_{n\epsilon Z}\eta_{1,2}^{\pm
(m,n)}e^{in\pi\sigma}e^{im\frac{2\pi}{t}\tau} \nonumber \\ 
\psi_{3,4}^{\pm}&=& \sum_{n\epsilon Z+\frac{1}{2}}\eta_{3,4}^{\pm
(m,n)}e^{in\pi\sigma}e^{im\frac{2\pi}{t}\tau} \\
Z_{1,2}^{\pm}&=& \sum_{n\epsilon Z} \alpha_{1,2}^{\pm(m,n)} cos 
(n\pi\sigma) e^{im\frac{2\pi}{t}\tau} \nonumber \\
Z_{3,4}^{\pm}&=& \sum_{n\epsilon Z+\frac{1}{2}} 
\alpha_{3,4}^{\pm(m,n)} cos (n\pi\sigma) e^{im\frac{2\pi}{t}\tau} 
\nonumber 
\end{eqnarray}
where it is understood that the sum over m always runs over the integers. 
For $\tilde{\psi}_k^{\pm}$ take simply $\sigma\rightarrow-\sigma$.

In contrast with the heterotic computation, we have here in general a 
t-dependence of the amplitude, meaning that non-BPS states are contributing
as well as BPS ones, making the evaluation of the determinants
not trivial at all. However, the infrared leading
singularity, given by the limit $t\rightarrow\infty$, only picks
up the $n=0$ contribution. In this limit the twisted fermions decouple (due 
to the lack of this zero mode) and the bosonic part gives
\be
\prod_{m=1}^{\infty}\left[1-\left(\frac{(\lambda+\lambda^{'})
\bar{\mu}t}{m\pi}\right)^{2}\right]^{-1}
\left[1-\left(\frac{(\lambda-\lambda^{'})
\bar{\mu}t}{m\pi}\right)^2\right]^{-1}=
\left[\frac{(\lambda+\lambda^{'})\bar{\mu}t}{\sin((\lambda+\lambda^{'})
\bar{\mu}t)}\right]
\left[\frac{(\lambda-\lambda^{'})\bar{\mu}t}
{\sin((\lambda-\lambda^{'})\bar{\mu}t)}\right]
\ee
The leading singular part for $\partial_{\mu}^2F_{g,n}$ will be then given
by the $\lambda^{2g}\lambda^{'2n}$ term in the expansion of
\be
\int_0^{\infty}\frac{dt}{t}
\left[\frac{(\lambda+\lambda^{'})\bar{\mu}t}{\sin((\lambda+\lambda^{'})
\bar{\mu}t)}\right]
\left[\frac{(\lambda-\lambda^{'})\bar{\mu}t}{\sin((\lambda-\lambda^{'})
\bar{\mu}t)}\right]e^{-|\mu|^{2}t}
\label{qed} \ee
where the term $e^{-|\mu|^{2}t}$ is due to the bosonic D-D zero mode
and it reproduces the same
singularity structure of the heterotic locus (\ref{c1}).
Is is worth while to mention that this gauge field, possible candidate
dual of the corresponding heterotic vector field, is in the same multiplet
of a D-D scalar.

\section{Conclusions}
We have studied in this paper a class of higher derivative F-terms of the 
form $F_{g,n}W^{2g}(\Pi f)^{n}$ (with W the gravitational superfield 
and $\Pi$ the chiral projector acting on a general
function $f$ of the complex vector superfield the heterotic dilaton 
belongs to) coming from the $N=2$ low-energy
lagrangian. It has been found that in the heterotic side, like the 
case for the $F_g$'s, only BPS states contribute to the amplitude, while
this is not what we found in the D-brane picture. It is important to stress
here that even if the gauge field we used in the D-brane computation is 
not the only one reproducing the heterotic result in the leading 
singularity limit, it does not exist a physical state giving 
the heterotic structure for the $F_{g,n}$ terms away from this limit.
It should be pointed out, however, that this result is not in 
contradiction with the conjectured heterotic-type II duality, because we 
compared two amplitudes computed perturbatively at 1-loop, but according
to different arrangements of the perturbative expansion. There is no reason
known, at least to us, for which we should expect an exact matching of 
the $F_{g,n}$ couplings. On the other hand, the leading behaviour
is the same, according to the given duality.

We think that a further study in this direction can lead to further
checks and help to clarify the nature of the $N=2$ heterotic-type II duality 
beyond the effective action; in particular it should be interesting
to compute similar couplings taking into account vector gauge fields of 
general moduli.  

An other point we believe should be attractive and studied better is the 
appearence of the free energy of the c=1 string at radius R, considering
also the role played by the self-dual radius in describing the coniflod 
singularity. It is interesting to note that, as in \cite{agnt}, 
eqs.(\ref{c1}) and (\ref{qed}) coincide with the effective action of QED 
\cite{sch} 
in a constant background  $F$ given by $\lambda=\sqrt{(F^2+F\tilde F)/2}$ 
and $\lambda^{'}=\sqrt{(F^2-F\tilde F)/2}$ !

{\bf{Acknowledgements}}

We would like to thank E. Gava and K.S. Narain for the enlightening
discussions from which the idea of this paper originated and for helpful 
suggestions.

\end{document}